# Realization of biferroic properties in
# $La_{0.6}Sr_{0.4}MnO_3$ / $0.7Pb(Mg_{1/3}Nb_{2/3})O_3 − 0.3(PbTiO_3)$ epitaxial superlattices

Ayan Roy Chaudhuri[a], R. Ranjith, and S. B. Krupanidhi[b]
*Materials Research Centre, Indian Institute of Science, Bangalore 560 012, India*

R.V.K. Mangalam, and A. Sundaresan
*Chemistry and Physics of Materials Unit, Jawaharlal Nehru Centre for Advanced Scientific Research, Jakkur, Bangalore 560 064, India*

S. Majumdar, and S.K. Ray
*Department of Physics and Meteorology, Indian Institute of Technology, Kharagpur721 302, India*

# Abstract:

A set of symmetric and asymmetric superlattices with ferromagnetic $La_{0.6}Sr_{0.4}MnO_3$ and ferroelectric $0.7Pb(Mg_{1/3}Nb_{2/3})O_3 − 0.3(PbTiO_3)$ as the constituting layers were fabricated on $LaNiO_3$ coated (100) oriented $LaAlO_3$ substrates using pulsed laser ablation. The crystallinity, magnetic and ferroelectric properties were studied for all the superlattices. All the superlattice structures exhibited a ferromagnetic behavior over a wide range of temperatures between 10K and 300K, whereas only the asymmetric superlattices exhibited a reasonably good ferroelectric behaviour. Strong influence of an applied magnetic field was observed on the ferroelectric properties of the asymmetric superlattices. Studies were conducted towards understanding the influence of conducting LSMO layers on the electrical responses of the heterostructures. The absence of ferroelectricity in the symmetric superlattice structures has been attributed to their high leakage characteristics. The effect of an applied magnetic field on the ferroelectric properties of the asymmetric superlattices indicated strong influence of the interfaces on the properties. The dominance of the interface on the dielectric response was confirmed by the observed Maxwell – Wagner type dielectric relaxation in these heterostructures.

---
[a] Electronic mail: ayan@mrc.iisc.ernet.in
[b] Author to whom correspondence should be addressed; FAX: +9180 2360 7316
   Electronic mail: sbk@mrc.iisc.ernet.in

# Introduction:

Fabrication and stabilization of new materials that do not occur naturally has been the interest of study in recent trends of materials research[1]. Artificial superlattices provide an opportunity to fabricate materials with carefully engineered properties. The control of desired properties can be achieved by tailoring the lattices[2]. Superlattices, which are composed of thin layers of two or more different structural counterparts stacked in a well defined sequence, may exhibit remarkably different properties that do not exist in either of their parent compounds. For example $(LaFeO_3/LaCrO_3)$ superlattices fabricated on a (111) oriented $SrTiO_3$ substrate exhibit a ferromagnetic behaviour, where either of the parent material is antiferromagnetic[3]. Similarly perovskite based superlattices have shown new or enhanced properties, such as high temp superconductivity, ferroelectricity, multiferroicity[4-10] etc. There are many factors which influence the superlattice (SL) properties, such as nature of the constituent materials, their thickness, surface morphology, interface structure between the sublayers, and the substrate. The lattice mismatch and the difference in thermal expansion coefficients among the substrate and the individual materials are the most important contributors, which in turn determine the electrical and/or magnetic properties of the superlattices[11].

Materials, in which magnetism and ferroelectricity coexist,[12] are termed as biferroics, which may possess coupling between the electrical and magnetic order parameters, resulting in magnetic properties that can be tuned by the application of electric field and vice-versa. Biferroic materials possessing this coupled magnetic and ferroelectric order parameters are also termed as multiferroics. These materials with mutually controllable ferromagnetism and ferroelectricity are potential candidates for novel device applications e.g. transducers and storage devices with parametric values and flexibility[13]. There are two open issues in this field of research: what is the origin of multiferroic nature of these materials and how to enhance these properties for suitable applications. Single phase materials with coexisting magnetic and ferroelectric ordering are very rare in number[13,14].The scarcity of naturally occurring or synthesized single phase multiferroic materials limits the opportunities to address the aforementioned issues. Study of materials possessing coupled magnetic and electrical order parameters has been revitalized[15] to design materials with induced novel multiferroic behaviour. Various approaches have been made to design and synthesize artificial structures that exhibit multiferroism. One of them is doping either a magnetic impurity in a ferroelectric host or a ferroelectric impurity in a magnetic host, or designing composites with ferroelectric and ferromagnetic hosts. The other approach consists of designing multilayers and superlattices with two or more distinct compounds,[16] where due to breaking of time reversal symmetry it may be possible to induce ferromagnetism and/or ferroelectricity. In order to observe good ferroelectric properties in case of multilayers and superlattices, the total structure should be electrically insulating, but ferromagnetic materials are commonly electrically conducting below Curie temperature. Thus the SL structures are to be engineered in such a way by optimizing the magnetic layer thickness and other process parameters that the entire structure should be insulating in nature. Recently there is an upsurge of interest in fabricating engineered epitaxial superlattices and multilayerd thin films consisting alternative layers of ferromagnetic and ferroelectric materials. Possible accurate control over crystallographic orientations, layer thicknesses and interfacial roughness in the epitaxial heterostructures make them attractive for precise magnetoelectric studies. There are reports on perovskite based artificial multiferroic superlattices and multilayers, where different manganates such as $Pr_{1-x}Ca_xMnO_3$ (PCMO), La$_{1-x}$Ca$_x$MnO$_3$ (LCMO), La$_{1-x}$Sr$_x$MnO$_3$ (LSMO), have been used as the ferromagnetic layers while BaTiO$_3$ (BTO), Ba$_{1-x}$Sr$_x$TiO$_3$ (BST) etc. have been used as the ferroelectric layers[7-10]. Most recently there was a theoretical prediction of magnetoelectric effect in Fe/ BaTiO3 multilayer[17]. In such artificial heterostructures containing alternative layers of a ferroelectric and a ferromagnetic material the magnetoelectric coupling can be mediated via elastic strain. Strain mediated coupling requires intimate contact between a piezomagnetic (or magnetostrictive) material and a piezoelectric (or electrostrictive) material, which can be achieved conveniently in epitaxial heterostructures.

In the present study a series of epitaxial superlattices composed of ferromagnetic LSMO (x=0.4) and ferroelectric/relaxor ferroelectric 0.7Pb(Mg$_{1/3}$Nb$_{2/3}$)O$_3$ − 0.3(PbTiO$_3$) (PMN-PT) have been fabricated on LaNiO$_3$ (LNO) coated (100) oriented LaAlO$_3$ (LAO) substrates using pulsed laser ablation. Electrostrictive PMN-PT having a very high piezoelectric coefficient (d$_{33}$ = 2000 pC/N) compared to other ferroelectric materials, such as PZT, BaTiO$_3$, BST etc. could give rise to a large room temperature magnetoelectric coupling with the magnetostrictive LSMO. Moreover the effect of magnetic layers on the ferroelectric and dielectric relaxation properties of a relaxor material can also be of scientific interest. A systematic study on the growth and biferroic properties of PMN-PT/ LSMO superlattices are dealt in the present paper. Two sets of superlattices with four different periodicities (6nm, 9nm, 13nm, and 16nm) were fabricated with constant total thickness of the heterostructures. The 1$^{st}$ set consisted of symmetric superlattices, with equal thicknesses of ferromagnetic and ferroelectric sublayers. In the 2$^{nd}$ set we fabricated asymmetric superlattices in which the magnetic sublayer thickness was kept constant at ~2nm and the ferroelectric sublayer thickness was varied.

## Experimental:

Phase pure ceramic targets of LNO, PMN-PT and LSMO were prepared using precursors supplied by Aldrich Chemicals (99.9%). Thin film capacitors were fabricated by pulsed laser deposition (PLD), using a 248 nm KrF excimer laser (Lambda Physik COMPex) on single crystal LAO substrates placed at a distance of 4 cm from the targets to get a fluence of 2.5 J/cm$^2$. The base pressure of the chamber was brought down to $1 \times 10^{-6}$ m.bar prior to each deposition. LNO bottom electrode was deposited under 300 m Torr. Oxygen pressure with a substrate temperature of $700^{\circ}$C and pulse repetition rate of 3 Hz. PMN-PT/ LSMO superlattices were then deposited under 100 m. Torr Oxygen pressure, keeping the substrate temperature at 700 $^{\circ}$C with pulse repetition rate of 3 Hz for LSMO layer and 5 Hz for PMN-PT layer. The films deposited were brought back to room temperature immediately after deposition with a cooling rate of 13 $^{\circ}$C min$^{-1}$. Two types of superlattices with varying periodicities (6 nm, 9 nm, 12nm, 16nm) were fabricated. Type 1: symmetric superlattices with equal thickness of the ferroelectric and the ferromagnetic material and Type 2: asymmetric superlattices with constant the magnetic layer thickness ($\sim$ 2 nm) and increasing the ferroelectric sublayer thickness. The total thickness of the superlattices were measured to be 250 $\pm$ 5 nm by cross sectional scanning electron microscopy (Quanta). Gold dots of area $1.96 \times 10^{-3}$ cm$^2$, deposited by thermal evaporation technique using a shadow mask were used as top electrodes.

The $\theta$-$2\theta$ and Phi ($\Phi$) scans were recorded on the samples for crystallographic and epitaxial characterizations using a Phillips X'Pert MRD Pro X Ray diffractometer (Cu $K_{\alpha}$ $\lambda = 0.15418$ nm). The dc magnetization measurements were performed using a vibrating sample magnetometer (VSM) by placing the sample in a physical properties measurement system (PPMS) (by Quantum design, USA) with the magnetic field parallel to the growth direction. A Radiant Technology Precision ferroelectric workstation was used to measure the room temperature ferroelectric polarization hysteresis at frequencies ranging from 200 Hz to 2 kHz. The leakage current characteristic was measured at room temperature using a Keithly 236 Source Measure Unit (SMU), and the dielectric properties were studied within a wide temperature range between $-120^{\circ}C$ to $400^{\circ}C$ using an Agilent 4294A impedance analyzer in the frequency range of $10^2 – 10^6$ Hz.

## Results and Discussion:

### a) Structural Characterization

Figure 1 (a) and (b) show a typical $\theta$ -$2\theta$ scan around the (100) fundamental peak (18-25° in 2θ) for the symmetric and asymmetric SL structures respectively with 16 nm periodicity. The denoted number $i$ indicates the $i^{th}$ satellite peak. The higher order satellite peaks in the X Ray Diffraction pattern adjacent to the main diffraction peak of the averaged lattice indicated that all the films were heterostructurally coherent with sharp interfaces. To investigate the in plane coherence of the superlattices, a Φ scan was performed around the (103) plane with fixing the $2$ value calculated from the (100) peak positions of the substrate and the superlattices. Four distinct peaks with 90° spacing clearly indicate the fourfold symmetry as expected for the cubic/tetragonal perovskite structures. The nature of the phi scans shown in figure 2, confirm that the superlattices are grown epitaxially "cube on cube". The enhanced width of the diffraction peaks shows that the films are under strain. The lattice mismatch (ε) between the film and the substrate can be calculated from the following relation:

$$\varepsilon = [(a_{substrate} - a_{film}) / a_{substrate}] \times 100\% \qquad (1)$$

Both LSMO (a = 3.87Å) and PMNPT (4.025Å) having respective lattice mismatches of -0.28% and -4.30% with the LNO (a= 3.859Å in bulk) experience compressive in plane stress at the bottom most layer of the superlattices. LNO also experiences a compressive stress due to -1.82% lattice mismatch with the LAO substrate (a=3.79Å). In effect the entire set of heterostructures experienced an in plane compressive stress and the stress increased with increasing periodicity of the superlattices. This was evident from the shift of the film peak positions towards lower $2\theta$ values with increasing periodicity. For symmetric set, the out of plane d-spacing of the averaged lattice ranges between 4.01 and 4.13Å, whereas in case of asymmetric superlattices it ranges between 4.12 and 4.19Å. This indicates that the strain is larger in case of asymmetric superlattices. Periodicities of the SL structures were calculated from the X- Ray diffraction pattern using the following equation[18]:

$$\Lambda = \frac{\lambda}{2(\sin\theta_{k+1} - \sin\theta_k)} \qquad (2)$$

Where, $\Lambda$ is the SL periodicity, $\lambda$ is the Cu K$_\alpha$ wavelength (0.15418 nm), $\theta_{k+1}$ and $\theta_k$ are two adjacent SL peaks in $\theta$-$2\theta$ geometry. The calculation from the XRD patterns was in good agreement with the thickness calibration carried out on individual layers, bilayers and multilayers for various number of laser pulses using a Scanning Electron Microscope (Quanta). Asymmetries present in the intensity distribution of the SL reflections have been studied extensively by static phonon approach and structure factor calculations[19], and are attributed to the fluctuation in the periodicity and strain of the system. This could be due to the annealing effect experienced by the initial layers of deposition until the top layers are completed, which is unavoidable. Moreover the second order term corresponding to the strain variation and the corresponding modification of structure factor introduces an asymmetry in the intensity distribution.

**B) Ferromagnetic characterization:**

In order to characterize the ferromagnetic properties of the superlattices, magnetic hysteresis (M-H) measurements were performed at different temperatures ranging from 10 K to 300K with a magnetic field swept between − 0.3T to + 0.3T. Figure 3 (a) and (b) show the M-H hysteresis loops of 13 nm symmetric and asymmetric SL at 10K and 300K. Well-defined coercivity exhibited by all the samples confirmed the ferromagnetic nature of the heterostructures in this measured range of temperature. To look into the magnetization behaviours of the superlattices in further details one SL from each set with highest values of magnetization (6nm for the symmetric set, and 16 nm for the asymmetric set) was selected and the remnant magnetization ($M_R$) values were compared with a 40 nm thick single layer LSMO film.

At lower temperature (10K) the $M_R$ values of the symmetric SL (17.29 emu/cm$^3$) and the asymmetric SL (34.13 emu/cm$^3$) were lower compared to that of the LSMO single layer film (54.45 emu/cm$^3$). At 300 K on the other hand the asymmetric SL showed the highest value of remnant magnetization (22.84 emu/cm$^3$). This is twice the value observed from the symmetric SL (11.57 emu/cm$^3$), and the single layer LSMO film (11.68 emu/cm$^3$), which have comparable values of remnant magnetization at this temperature.

In order to explain these observations, let us consider first the architecture of the heterostructures. In case of the symmetric superlattices, the total thickness of LSMO (80nm) layers in a heterostructure is the same for all the periodicities. On the other hand the individual LSMO layers for the asymmetric superlattices remained constant (2nm) in each period. Therefore, depending upon the SL periodicity, the total thickness of the LSMO varied between 20 nm (for highest periodicity) to 40 nm (for lowest periodicity). In figure 4, the $M_R$ and coercive field ($H_C$) at 300 K have been plotted as a function of SL periodicity for both symmetric and asymmetric sets. Both the $M_R$ and saturation magnetization ($M_S$) values of the symmetric superlattices decreased with increasing periodicity, while the $H_C$ increased. The dimensional dependence of the coercivity could be attributed to the presence of magnetically "dead" layers at the between the LSMO and PMNPT interfaces.[20]

The asymmetric superlattices on the other hand showed increasing $M_R$ and decreasing $H_C$ with increasing SL periodicity, in contrary to the properties observed for the symmetric set. The $M_R$ values of the asymmetric superlattices are almost independent of temperature from 10 K to 300K. These observations can be explained by taking into consideration the effect of strain in the heterostructures. The magnetic properties of thin film Manganites are highly susceptible to lattice strain. Strain can change the easy direction of magnetization in thin films.[21,22] It has been reported that in thin films of manganites, the easy direction of magnetization is always along the direction of largest cell parameters.[23]

Lattice strain can also change the co-operative Jahn-Teller distortion present in the rare earth Manganties, which induces A-type antiferromagnetism in these materials. The increase of biaxial compressive strain gives rise to a tetragonal distortion, which on one hand reduces the in-plane Mn-O bond length and thus enhances the double exchange (DE) interaction. In addition to this recent theoretical and experimental investigations have shown that the electron-lattice interaction in thin films is very sensitive to the biaxial strain.[24-26] A compressive strain can increase the "bare" electron hopping amplitude, and thereby reduces the relative strength of the Jahn-Teller electron-lattice interaction. This can enhance the DE ferromagnetism in the thin film manganites[26]. In the present case, since all the superlattices are experiencing in-plane compressive strain, their enhanced ferromagnetism could be primarily attributed to the lattice strain driven reduction in co-operative Jahn-Teller distortion.

However, the current data is only qualitative and insufficient to comment on the exact mechanism behind the observed magnetic behaviours. In order to study the dimensional dependence and role of interfaces in the observed characteristics, temperature dependent magnetic and resistivity measurements, X-Ray reflectivity, and polarized neutron diffraction studies are currently under progress.

**C) Ferroelectric characterizations:**

Figure 5 (a) and (b) show the capacitance vs. voltage ($C$-$V$) curves for the 16 nm periodic symmetric and asymmetric superlattices respectively. The capacitance and dielectric loss factor of the asymmetric SL showed well defined butterfly loops expected from a ferroelectric material. The symmetric SL on the other hand, exhibited a non-linear dielectric response as a function of dc bias, with very slim loop. The dielectric loss factor did not show the nature expected from a ferroelectric material. Since this kind of $C$-$V$ loops of this nature can arise either from the change in the dielectric constant of the material or changes in interface depletion width.[27] Therefore observation of such loops is not sufficient to confirm the ferroelectric nature of the material. Polarization hysteresis ($P$-$E$) loops of the superlattices were thus measured at different frequencies to further investigate the ferroelectric nature of the heterostructures. Figure 6(a) shows room temperature (RT) $P$-$E$ loops of an asymmetric SL with 16 nm periodicity at different probing frequencies ranging between 200 Hz to 2 kHz. Frequency independent and saturated natures of the *P-E* loops indicate the intrinsic ferroelectric nature of the SL. Under an applied ac signal all the asymmetric superlattices showed saturated hysteresis loops (figure 6(b)), which clearly establishes their intrinsic ferroelectric nature. Remnant polarization ($P_r$) and coercive field ($E_c$) of the asymmetric set of superlattices were plotted against the SL periodicity in figure 7(b). They showed a monotonous decreasing trend with reducing PMN-PT thickness, which could be attributed to the increased conductivity of the heterostructures with decreasing dielectric thickness. Moreover the existence of dielectric passive layers at the interfaces can also degrade the ferroelectric properties with decreasing ferroelectric layer thickness.[28] All the *P-E* loops observed in this case were asymmetrically shifted towards the positive electric field axis. This has been discussed in detail elsewhere and attributed to the presence of in built depolarizing fields present in these heterostructures.[29]

Figure 8 shows the RT *P-E* loops observed from the symmetric superlattices. The shapes of the loops observed for all the four symmetric SL structures indicated that the heterostructures are electrically conducting in nature. A ferroelectric loop tester measures the switched charge *Q* of a sample, which is related to the polarization *P* for an ideal ferroelectric insulator by the equation:

$$Q = 2P_r A \qquad (4)$$

Where, $P_r$ is the remnant polarization, and *A* is the electrode area for a parallel plate capacitor. For a sample with finite conductivity the equation takes the following form.

$$Q = 2P_r A + \sigma E_a t \qquad (5)$$

Where $\sigma$ is the electrical conductivity, $E_a$ is the applied electric field, and *t* is the measuring time[27]. In our case the nature of the PE loops obtained from the symmetric superlattices correspond to a resistive leakage loop, where the polarization P can be expressed as

$$P = (\sum_{n=0}^{k} n\Delta V / R \times \Delta t) / A \qquad (6)$$

Where, $P$ is the polarization, $R$ is the resistance of the material, $\Delta t$ is the time step per point, $\Delta V$ is the voltage step of wave, and $n$ is the point number of digitized triangle wave[30]. Room temperature resistance measurement of the symmetric superlattices showed that the sample resistances lie in the range of $10^3$ ohm at an applied DC bias of 0.1 V, which establishes their conducting nature. Hence, the loops obtained from the P-E measurements can be considered as an artifact and not the intrinsic ferroelectric property of the materials. Loss of ferroelectricity due to higher conductivity of the symmetric SL rules out the possibility of any multiferroic coupling in these heterostructures.

Conducting nature of the symmetric superlattices has been further investigated by room temperature leakage current ($I$-$V$) measurement under a step voltage of 0.1 V and a delay time of 100ms. That the responses observed in this time frame was true leakage behaviour of the samples was confirmed by measuring leakage current versus time ($I$-$t$) response of the heterostructures at a constant DC bias of 1 Volt. Figure 9 shows the $I$-$t$ responses observed from one of the symmetric and one asymmetric SL. Since there is practically no change in the order of magnitude of the $I$-$t$ response within a time frame of 0.1s to 100s, it can be concluded that the time domain used to measure the $I$-$V$ response was beyond the transient regime and the $I$-$V$ responses show the true leakage behaviour of the samples.   Room temperature leakage current densities ($J$) of the symmetric and asymmetric set of SL have been plotted as a function of electric field ($E$) in figures10 (a) and 10 (b). Leakage current densities of the symmetric superlattices were of the order of $\sim 10^{-1}$ Amp/cm$^2$ at an applied 80 kV/cm DC electric field. This value is three to five orders of magnitude higher compared to that of the asymmetric superlattices in which depending upon SL periodicity the leakage current densities fall in the range of $\sim 10^{-4}$ to $\sim 10^{-6}$ Amp/cm$^2$ at the same applied field. This indicates the influence of the conducting LSMO layer on the electrical properties of the superlattices. In case of symmetric superlattices the leakage current density increased with increasing periodicity. On the contrary, for asymmetric superlattices it decreased with increasing periodicity. Constant thickness of LSMO sublayers combined with increase of insulating PMN-PT sublayer thickness enhances the leakage barrier for the asymmetric superlattices with increasing periodicity. This in effect increases the insulating characteristics of the entire heterostructure and could be responsible for the improvement of their ferroelectric properties.

Co-existence of ferroelectricity and ferromagnetism in the asymmetric set of superlattices essentially proves their biferroic nature. Two investigate the possible coupling between these two order parameters it is essential to study the effect of a magnetic field on the ferroelectric properties of the samples and vice-versa.

**D) Magnetic field effect on ferroelectric properties:**

In order to study the influence of a magnetic field on the ferroelectric properties of these biferroic heterostructures, the asymmetric superlattices were subjected to a magnetic field of 1.2T at room temperature with the magnetic field parallel to the growth direction. P-E hysteresis loops measured immediately after exposing the films to the magnetic field showed reduced asymmetry along the electric field axis accompanied by an increase in remnant polarization and coercive field values. Similar shift in P-E loops was observed on poling the samples with a 500 kV/cm DC bias, which has been reported elsewhere[29]. The observed effect could either be due to effective coupling between the residual magnetic moment in the SL and the applied electric field or it may be purely due to the effect of interfaces between the metallic FM and insulating FE layers. The magnetic field effect on the P-E loop has been represented in terms of difference in the coercive field ($\Delta 2E_C$) before and after exposing the films to the magnetic field: $\left| \Delta 2E_C \right| = \left| 2E_C(0) - 2E_C(H) \right|$, where $2E_C(0)$ is coercive field of the P-E loops measured before applying any magnetic field and $2E_C(H)$ is coercive field of the P-E loops measured after exposing the films to a magnetic field. Figure 11 shows that the $\Delta E_C$ increased with increasing SL periodicity. This could be attributed to the increase of lattice strain with increase of SL periodicity. The lattice mismatch between PMN-PT and LSMO coupled with discontinuous polarization at the interfaces could collectively give rise to a strong interfacial depolarizing field. On application of a magnetic field the magnetic domains get oriented along the filed directions and consequently alter the charges at the interfaces eventually leading to an alteration of the electric field at the interfaces. Even on removal of the magnetic field some of the domains get clamped and these charge accumulation at the PMN-PT/LSMO interfaces results in shifting the P-E hysteresis loops along the electric field axis. In the present case the magnetically tuned ferroelectric behaviour of the heterostructures indicates the possibility of an interface dominated phenomenon.

There is a debate on whether the magnetoelectric properties and the enhanced dielectric properties of the biferroic superlattices are bulk dominated or interface dominated[31]. Role of interfaces in dielectric property enhancement has been observed in different epitaxial ferroelectric thin films[32-34]. In the present case observed effect of a magnetic field on the ferroelectric responses of the heterostructures indicate a strong influence of the interfaces on these properties, which is not unexpected from the superlattice structures with a large number of interfaces involved. It is therefore necessary to investigate the role of interfaces and their dimensional dependence in tuning the SL properties. Moreover the coexistence of ferromagnetism and ferroelectricity at room temperature and the strong influence of a magnetic field on the P-E loops indicate the possibility of substantial magnetoelectric coupling in the superlattices. Magnetoelectric studies of these superlattices with varying magnetic field and temperature are currently under progress.

**E) Dielectric characterization:**

In order to study the influence of interface in the observed properties of the asymmetric superlattices, dielectric measurements were performed in a wide range of temperature and frequency. In many cases dielectric behaviour of multilayers and superlattices has been found to be dominated by the interfaces,[32] which has been explained in terms of Maxwell-Wagner (MW) effect. In this treatment the different sublayers are considered to have finite conductivity, and the overall polarization effects are dominated by charge accumulation at discontinuous interfaces within the heterostructures. The real ($\varepsilon'$) and imaginary ($\varepsilon''$) parts of the relative permittivity for such multilayered structures can be expressed as

$$\varepsilon'(\omega) = \frac{1}{C_0(R_i + R_b)} \frac{\tau_i + \tau_b - \tau + \omega^2 \tau_i \tau_b \tau}{1 + \omega^2 \tau^2} \tag{6}$$

and,

$$\varepsilon''(\omega) = \frac{1}{\omega C_0(R_i + R_b)} \frac{1 - \omega^2 \tau_1 \tau_2 + \omega^2 \tau(\tau_1 + \tau_2)}{1 + \omega^2 \tau^2} \tag{7}$$

where, $\tau_i$, $\tau_b$, and $\tau$ are the relaxation times of interface, bulk PMNPT and the entire dielectric multilayer respectively. $C_0$ is the geometric capacitance, $R_i$ and $R_b$ are resistances of the interface and dielectric PMNPT respectively and $\omega$ is frequency. The real part of dielectric constant has the same expression for Debye and MW type of relaxations, and therefore analysis of $\varepsilon'$ does not particularly give any useful information about the interfacial polarization effects.

On the other hand the imaginary permittivity does distinguish MW from Debye behaviour. In particular $\varepsilon'' \rightarrow 0$ as $\omega \rightarrow 0$ in a Debye system, whereas in a MW system $\varepsilon'' \rightarrow \infty$. By considering behaviour at zero and infinite frequency, equivalent expression to Eq. 7 can be given in terms of $\varepsilon_0$ and $\varepsilon_\infty$

$$\varepsilon''(\omega) = \frac{1}{\omega C_0 (R_i + R_b)} + \frac{(\varepsilon_0 - \varepsilon_\infty)\omega\tau}{1 + \omega^2\tau^2} \qquad (8)$$

Figure 12 shows the room temperature $\varepsilon''$ vs. frequency for all the asymmetric Superlattices. It has been observed that all the SL showed increasing loss at lower frequency regime which could possibly be due to the high leakage characteristic of the sample. Further increase of the dielectric loss at lower frequencies with increasing temperature strongly suggests the influence of conduction. The frequency response of the superlattices in the present case can be directly modeled by a MW expression, and a Debye expression cannot account for the low frequency behaviour. Parameters extracted from these MW fits at 300K have been summarized in table 1. The frequency dependence can be accounted for as follows: at high frequencies the mobile charge carriers (e.g. oxygen vacancy) with higher relaxation times cannot respond to the applied field, so that the measured capacitance is simply that of the insulating capacitors in series. At low frequencies on the other hand the charge carriers on the resistive layers do respond, so that most of the field drops across the layer with higher resistivity and thus the apparent decrease in dielectric thickness result in an increased capacitance. The enhanced conduction in the superlattices due to participation of the slow moving mobile charge carriers increases the dielectric loss at low frequencies. The low frequency dispersion of dielectric loss is more pronounced for the SL with 6 nm periodicity, which can be attributed to low thickness of PMNPT sublayer.

The onset of MW behaviour for the SL structures suggests that the observed modification of the PE hysteresis loops on application of a magnetic field arises due to the interfacial polarization mismatch and the effect of depolarizing field. The defect zones are formed at the interfaces between successive film layers which may have increased conductivity due to factors such as oxygen vacancy. An increased interface density would lead to a greater volume fraction of defect zones and the dielectric properties are thus strongly influenced by carrier mobility.

Studies are in progress in order to investigate the in depth dielectric relaxation dynamics and conduction behaviours of the superlattices as a function of temperature.

## Conclusions:

To summarize, symmetric and asymmetric superlattices of (PMN-PT/LSMO) have been successfully grown epitaxially on (100)-oriented $LaAlO_3$ substrate by pulsed laser ablation process. Despite the lattice mismatch between the $LaNiO_3$ coated substrate and LSMO (-0.28%) and PMN-PT (-4.30%), the films were grown heteroepitaxially. Both sets of superlattices showed good ferromagnetic characteristics in wide temperature range between 10K and 300K. The symmetric superlattices did not show any ferroelectric property, which is attributed to their leakage characteristics. On the other hand asymmetric superlattices with constant LSMO layer thickness ($\sim$ 2nm) showed good intrinsic ferroelectric nature, which was established conclusively by their polarization hysteresis and capacitance vs. voltage measurements. Coexistence of ferromagnetism and ferroelectricity at room temperature essentially proves the biferroic nature of the superlattices.

It has been concluded that the conducting LSMO layer was responsible for the leaky nature of the symmetric set of superlattices. In case of asymmetric set with constant LSMO thickness the leakage characteristics reduced with increasing thickness of PMN-PT spacer layer. The polarization hysteresis loops of the asymmetric superlattices were shifted asymmetrically along the electric field axis. Hysteresis loops measured after exposing these asymmetric superlattices to a magnetic field showed reduced asymmetry accompanied by increased values of remnant polarization and coercive field. The observed behaviour suggested strong influence of interfaces on the dielectric properties of the Superlattices. To verify the influence of interfaces dielectric relaxation behaviour of the SL were studied over a wide range of temperature and frequency. The dielectric loss data could directly be fitted to MW expression. The fitted data showed that conductivity of the films increases with decreasing PMN-PT sublayer thickness. The low frequency dispersion of the dielectric loss increased with increasing temperature for all the superlattices. This indicates the effect of conductivity on the dielectric properties. It is therefore suggested that in the present case the observed tuning of ferroelectric behaviour of the asymmetric SL after exposure to a magnetic field could be attributed to interfacial effects and might be indicative of induced magnetoelectric coupling.

| SL periodicity (nm) | $R_i + R_b$ (ohm) | $\varepsilon_0 - \varepsilon_\infty$ | $\tau$ (sec) |
|---|---|---|---|
| 6 | 2522.46 | 470.4484 | $1.18e^{-5}$ |
| 9 | 6500.9 | 484.87 | $1.20e^{-5}$ |
| 12 | 10000 | 510.86 | $1.23e^{-5}$ |
| 16 | 95000 | 630.28 | $1.37e^{-5}$ |

Table 1: Parameters obtained from Maxwell-Wagner fit of the room temperature $\varepsilon$" vs frequency curves of the asymmetric superlattices with different periodicities.

**Figure Captions:**

Figure1. X-Ray diffraction patterns of (a) symmetric and (b) asymmetric LSMO/ PMN-PT superlattice with 13 nm periodicity.

Figure2. Φ scan of a LAO substrate and a symmetric and an asymmetric superlattice.

Figure3. M-H hysteresis loops of (a) symmetric and (b) asymmetric PMN-PT/LSMO superlattices at different temperatures.

Figure4. Remnant magnetization ($M_R$) and coercive field ($H_C$) values of symmetric and asymmetric superlattices as a function of SL periodicity.

Figure5. C-V characteristics of 16 nm periodic (a) symmetric and (b) asymmetric superlattices.

Figure6. Room temperature P-E hysteresis loops of (a) asymmetric superlattice with 16 nm periodicity at different probing frequencies, (b) asymmetric superlattices of different periodicities.

Figure7. $P_S$ and $E_C$ values of the asymmetric superlattices as a function of periodicity.

Figure8. Room temperature P-E hysteresis loops of the symmetric superlattices of different periodicities.

Figure9. Room temperature *I-t* response of a symmetric and an asymmetric superlattice.

Figure10. Room temperature leakage current density vs. applied electric field for (a) the symmetric and (b) the asymmetric superlattices with different periodicities.

Figure11. $|\Delta 2E_C|$ vs. periodicities of the asymmetric superlattices.

Figure12. Room temperature ε" vs. frequency plot for asymmetric superlattices with different periodicities.

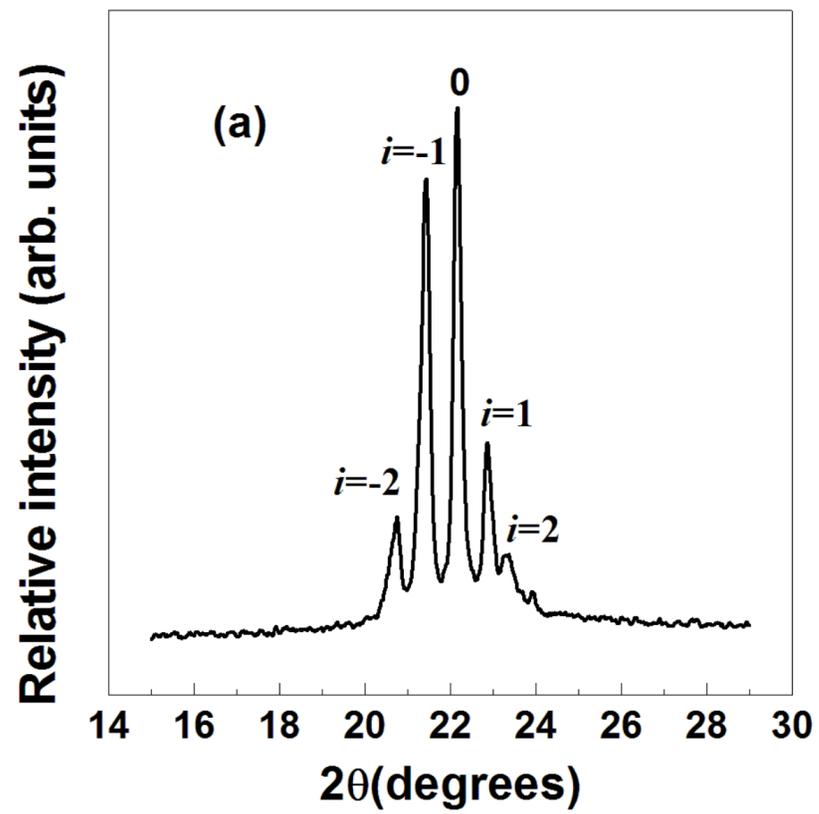

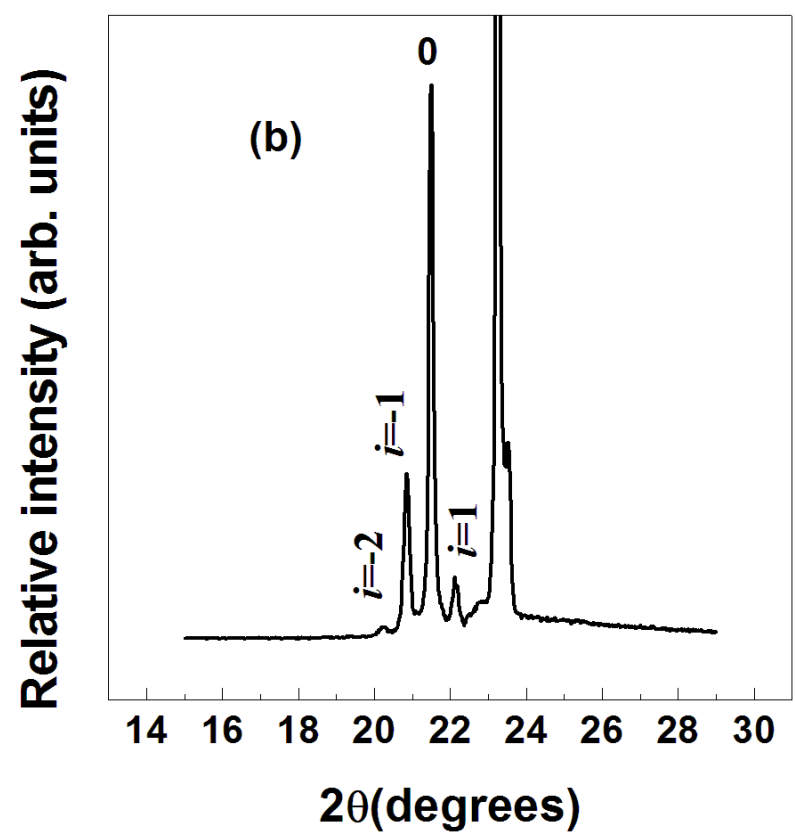

**Figure1**

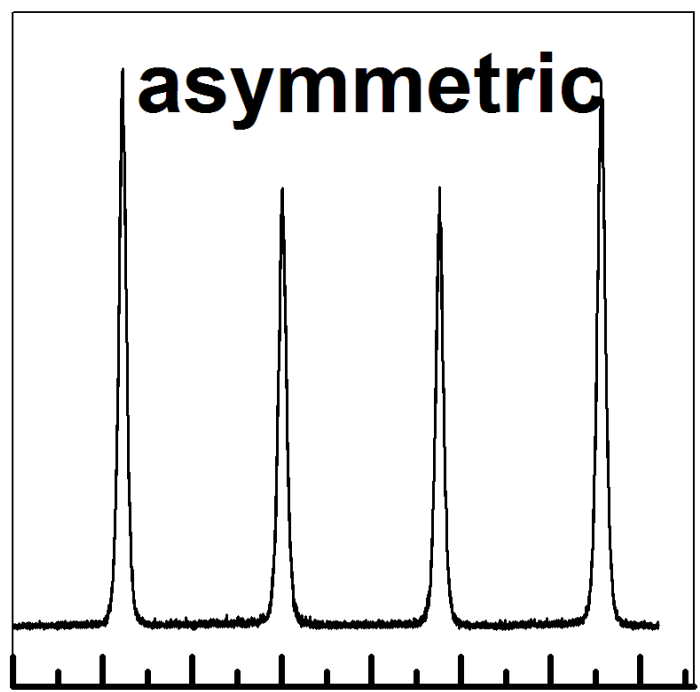

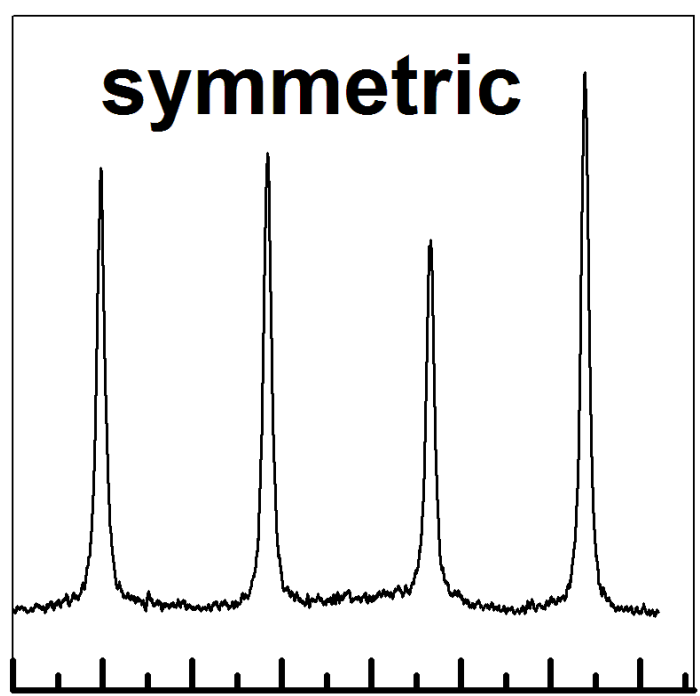

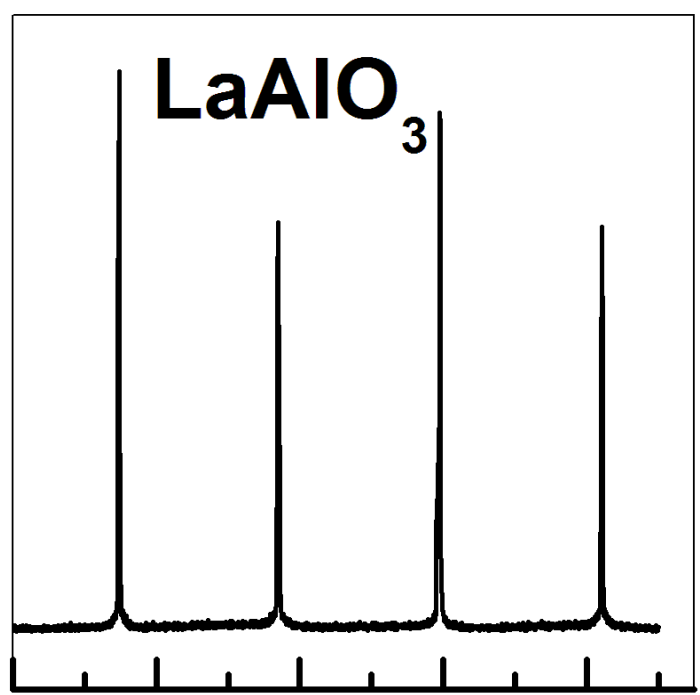

**Relative Intensity (Arb. Units)**

asymmetric

symmetric

LaAlO$_3$

$\Phi$ (degrees)

0    80    160    240    320

Figure 2

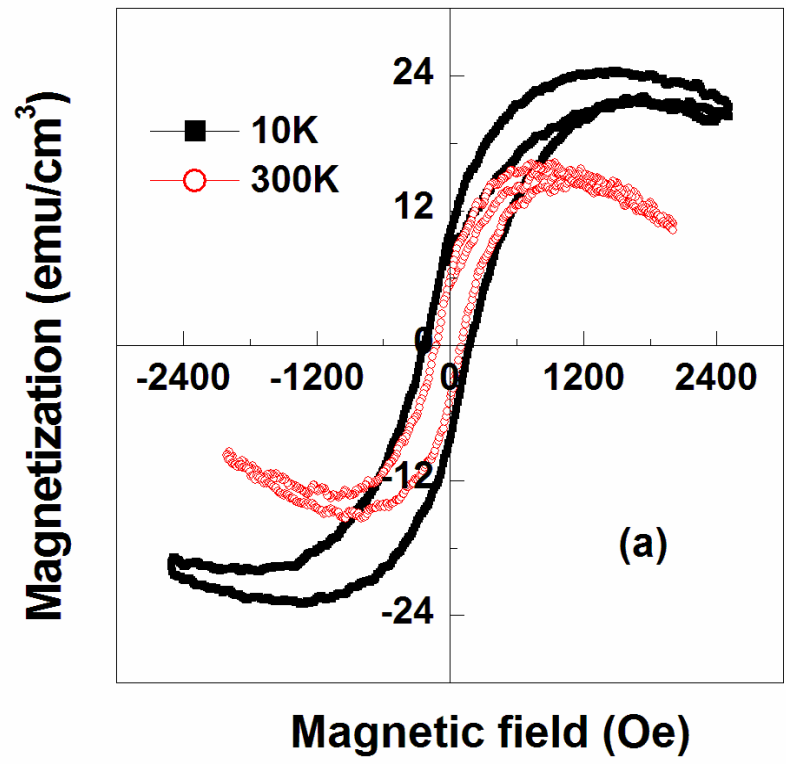

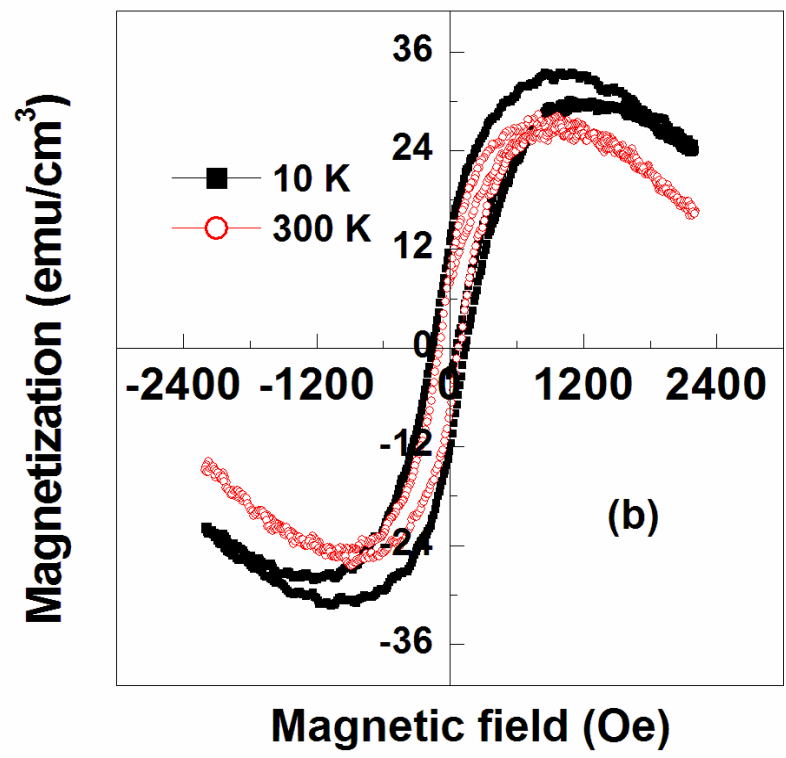

**Figure 3.**

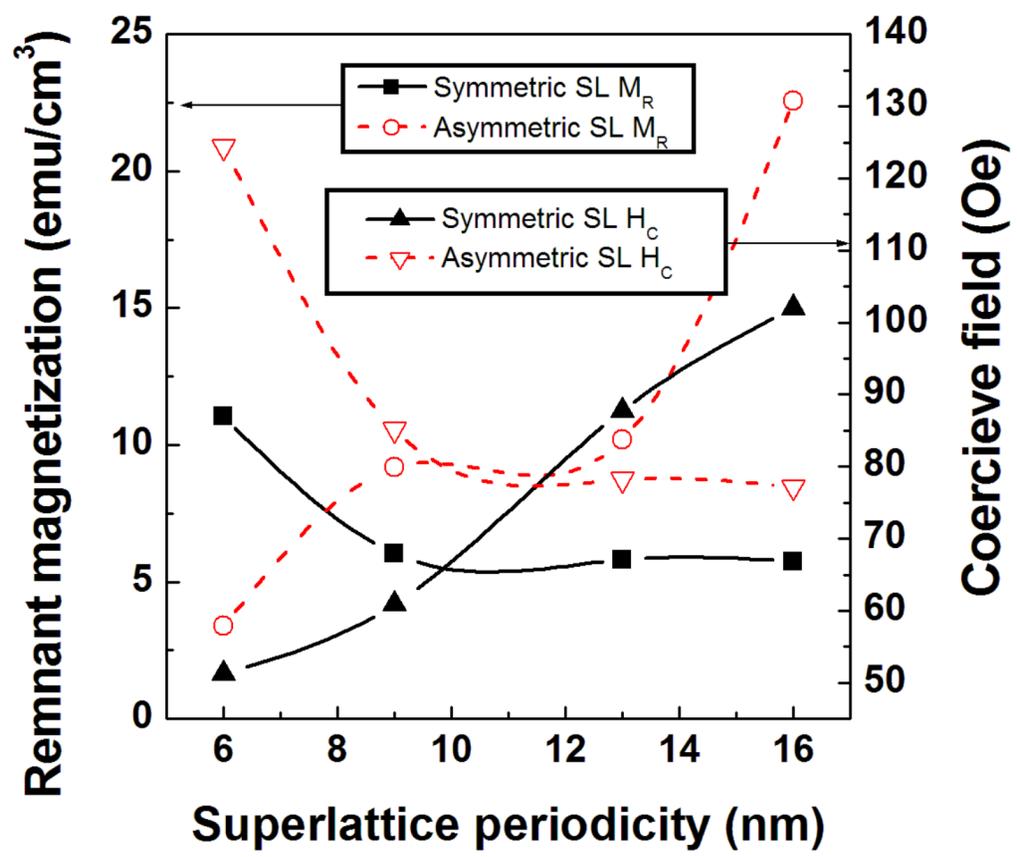

**Figure 4.**

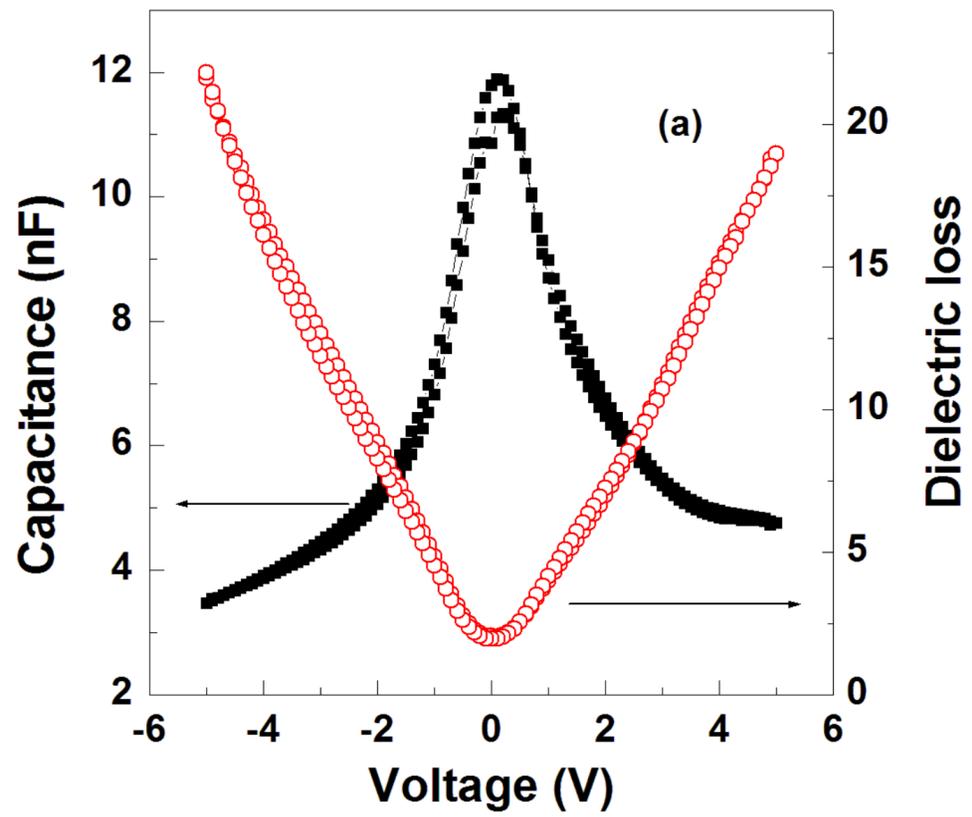

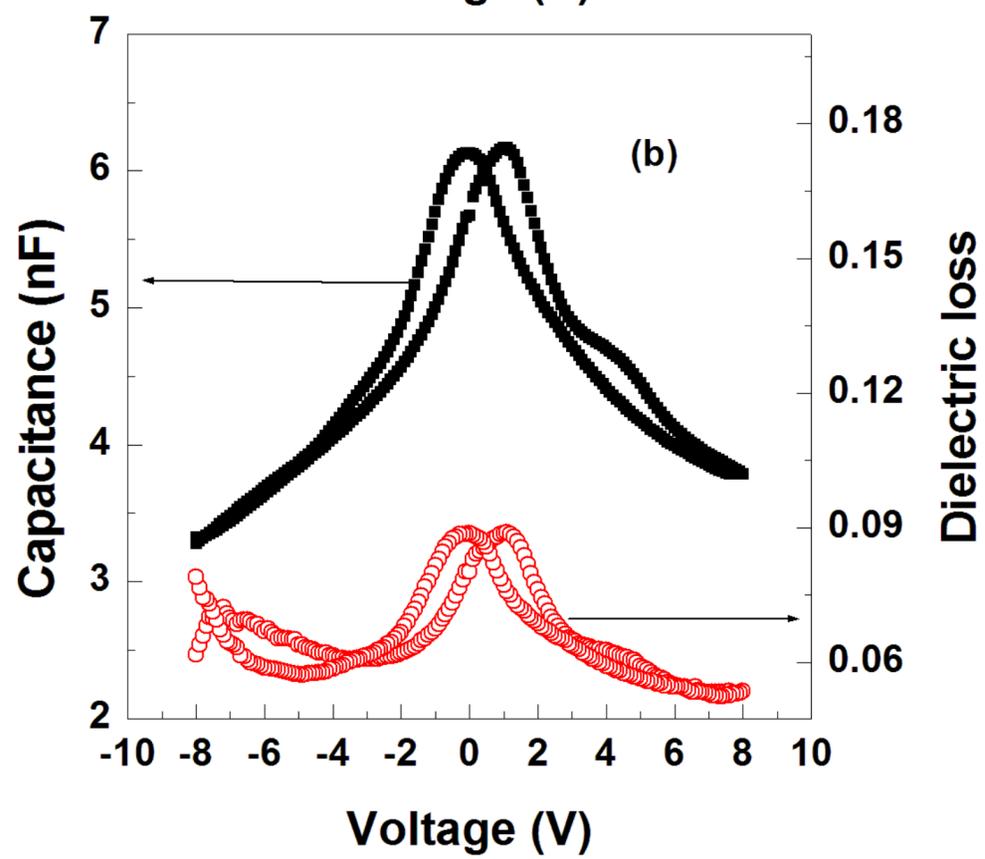

**Figure 5**

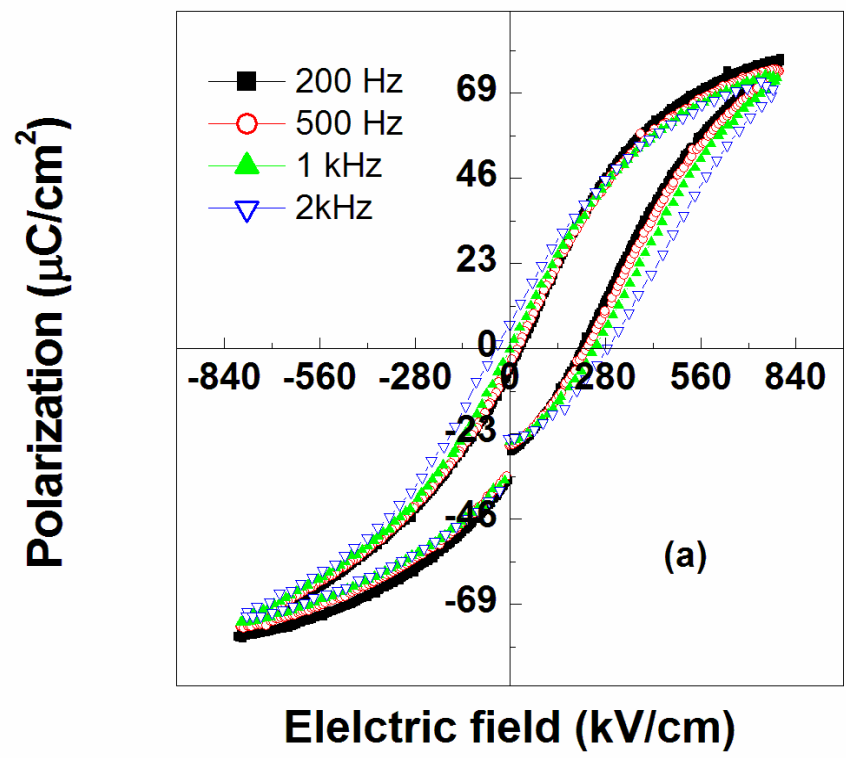

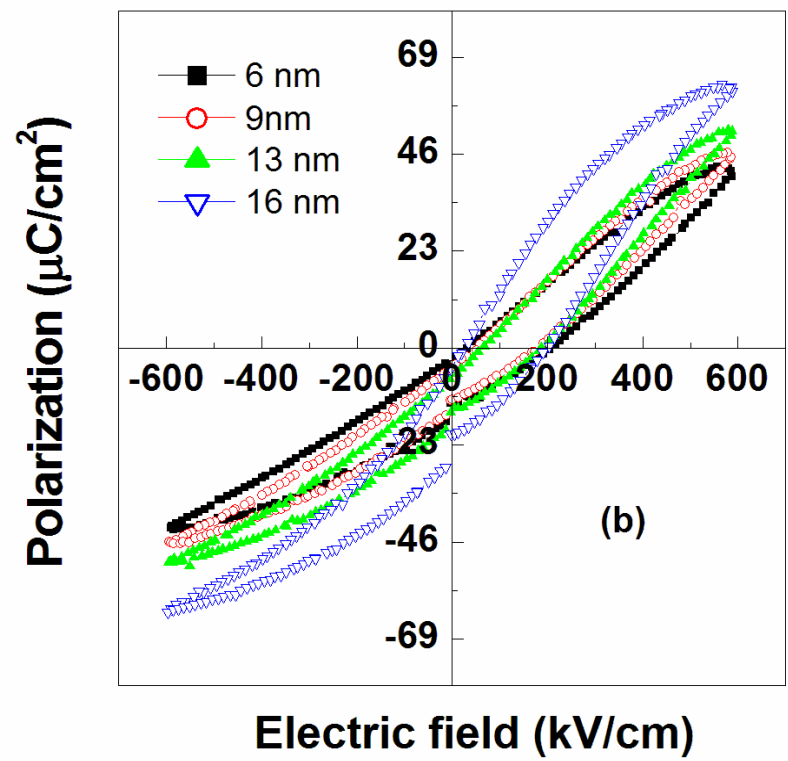

**Figure 6**

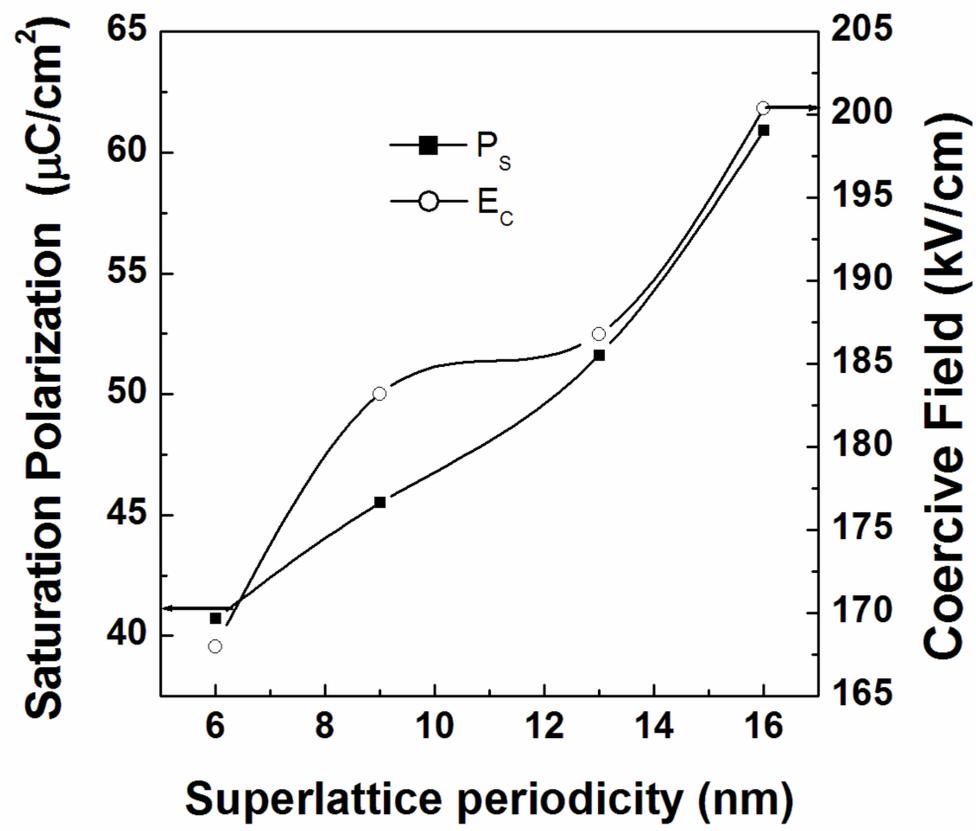

**Figure 7**

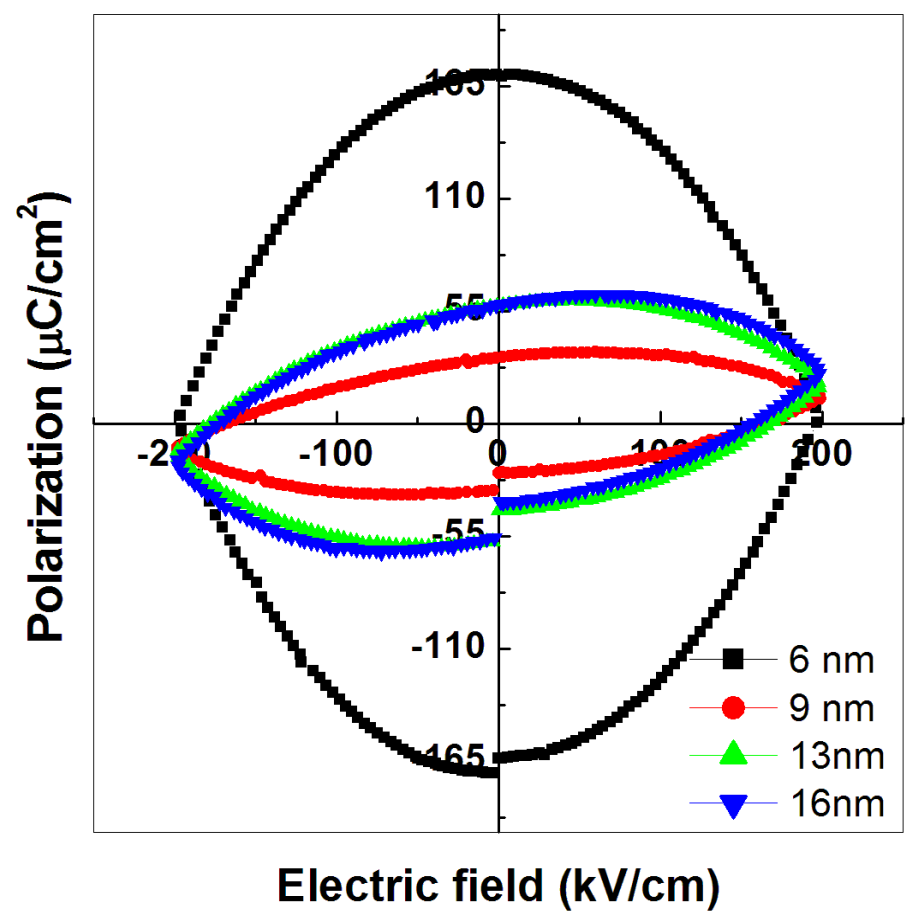

**Figure 8**

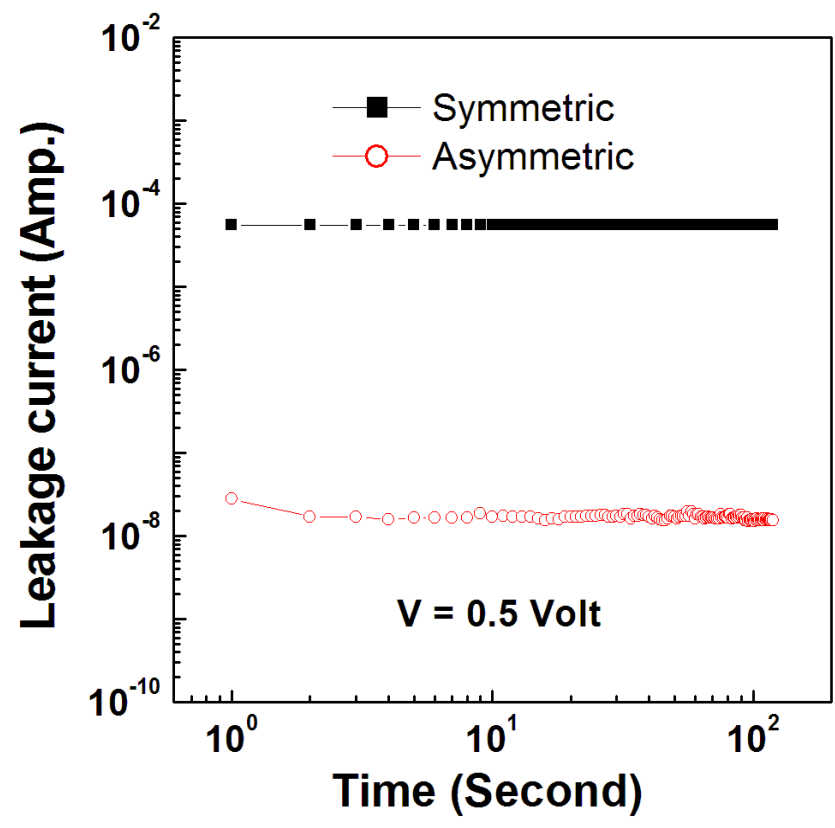

**Figure 9**

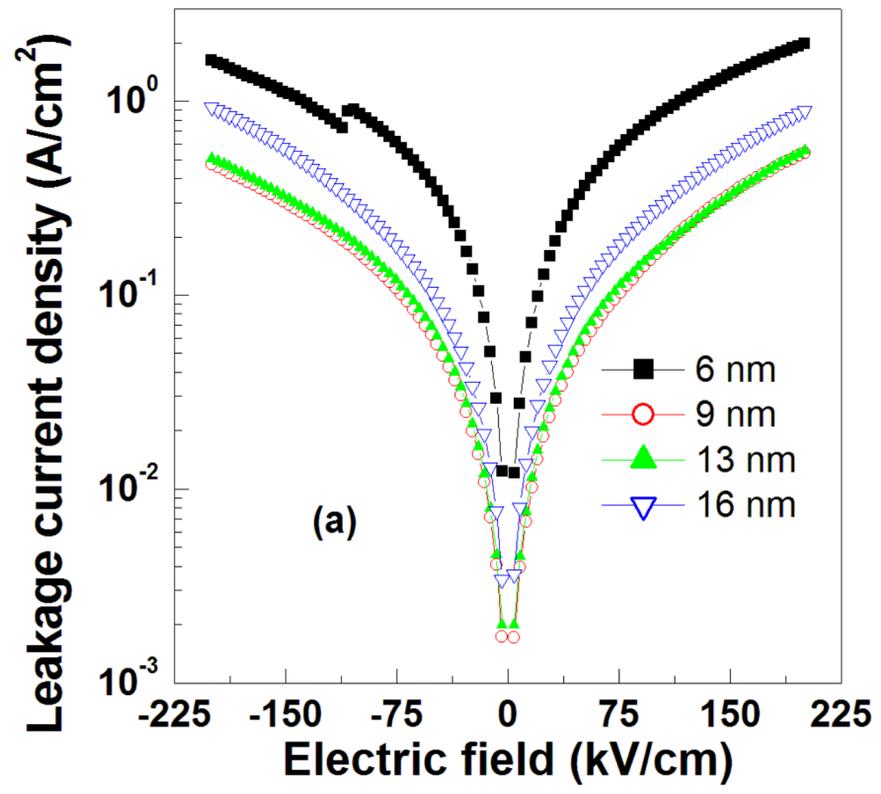

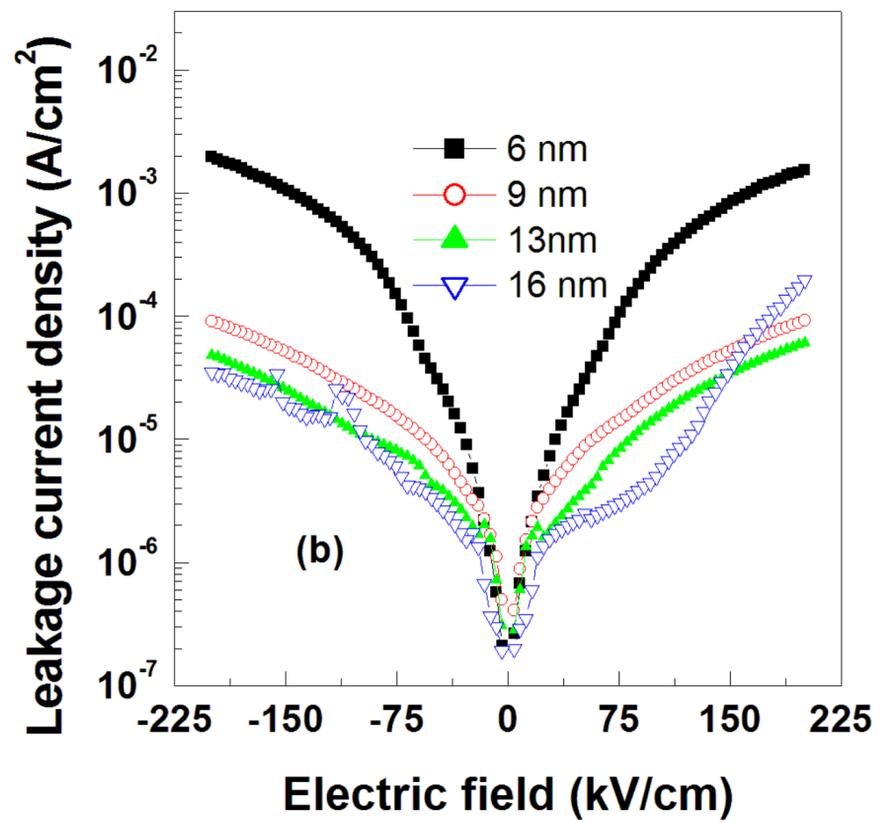

**Figure 10**

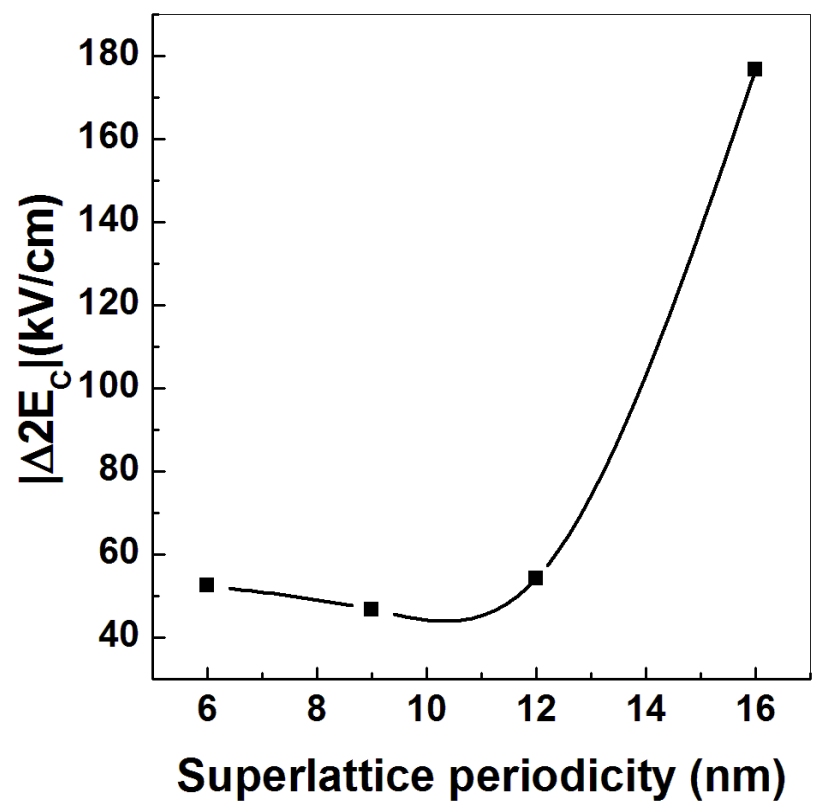

**Figure 11**

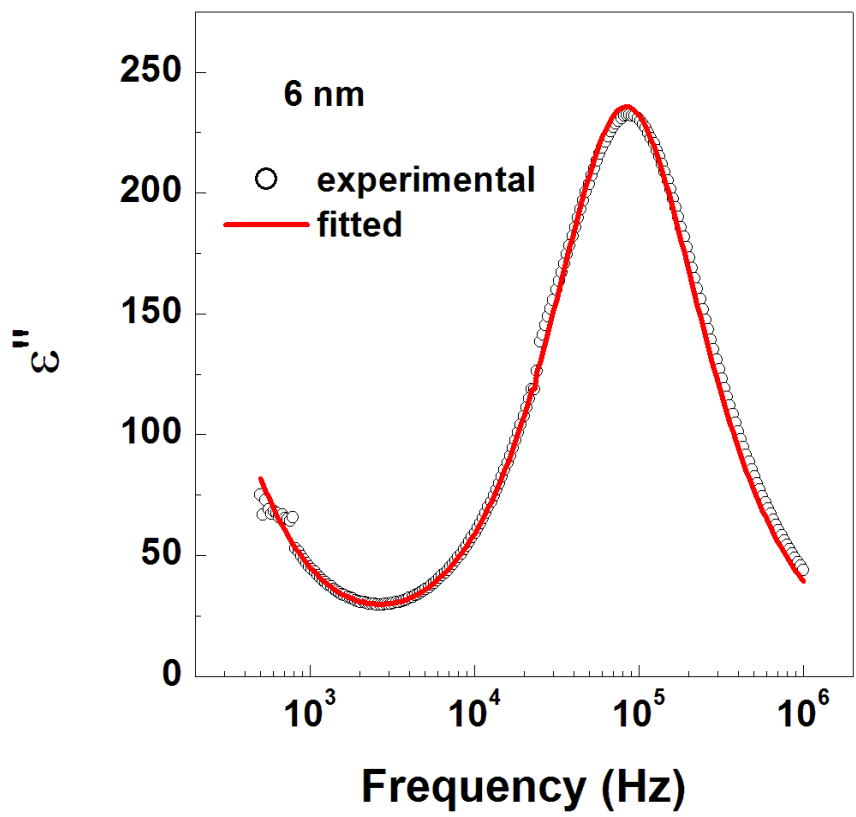

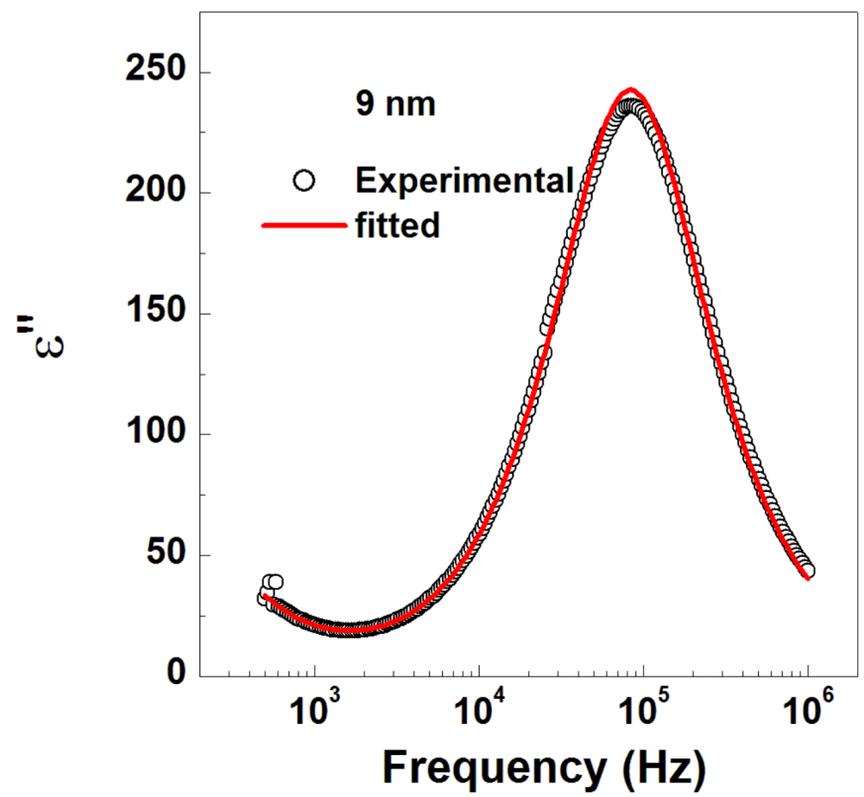

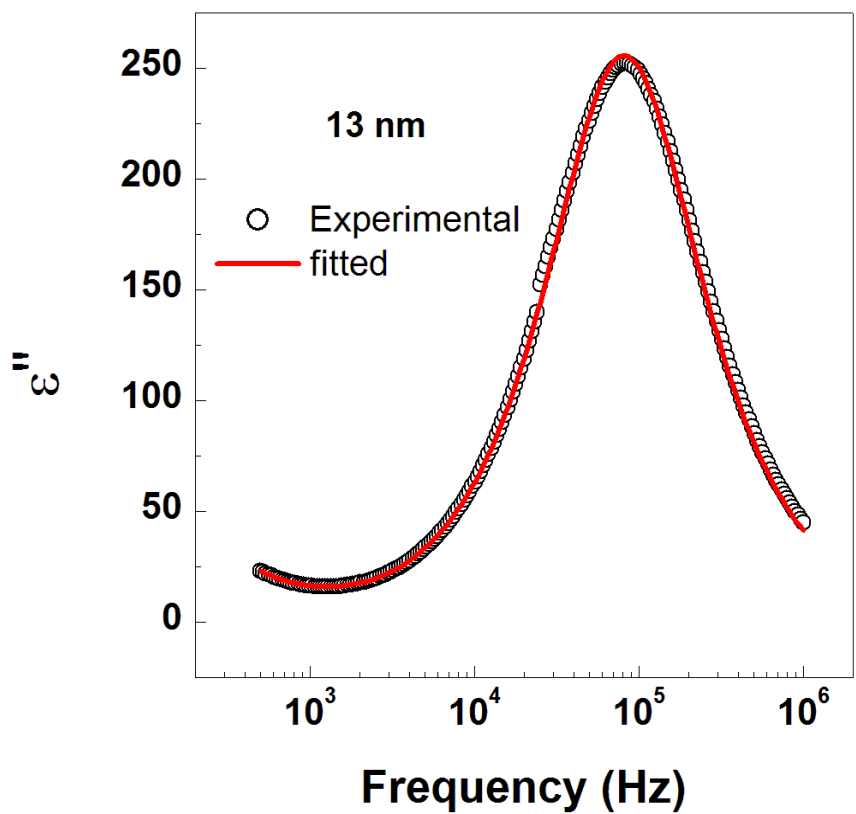

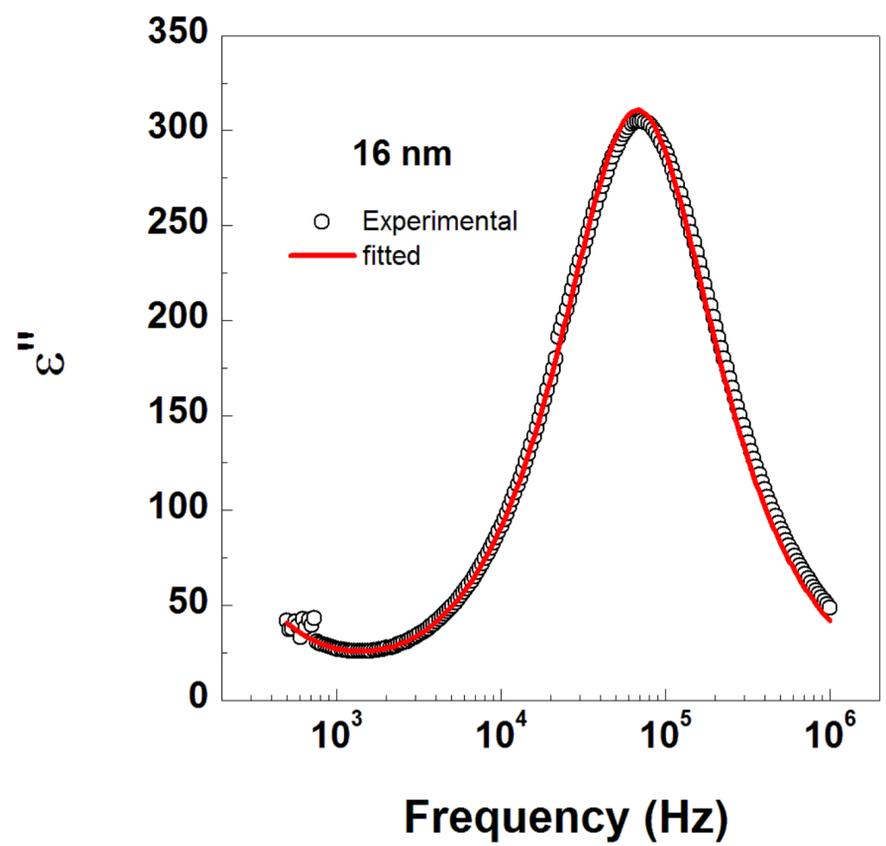

# Figure 12